\documentclass[draft]{article}
\def\today{26.7.13} 
\usepackage{amsmath,amsfonts,amsthm,amssymb,amscd}
\theoremstyle{plain} 
\newtheorem{theorem}{Theorem}[section]
\newtheorem{lemma}[theorem]{Lemma}

\newtheorem{definition}[theorem]{Definition}
\newtheorem{remark}[theorem]{Remark}

\newcommand{\R}{{\mathbb R}}

\newcommand{\Z}{{\mathbb Z}}

\newcommand{\e}{{\rm e}}

\newcommand{\I}{{\mathcal I}}

\newcommand{\Sc}{{\mathcal S}}
\newcommand{\Tc}{{\mathcal T}}

\newcommand{\C}{\mathbb{C}}

\def\supp{{\rm supp}}

\def\uno{{\kern+.3em {\rm 1} \kern -.22em {\rm l}}}

\def\norma#1{\left\| #1\right\|}

\def\mod#1{\left\lfloor#1\right\rceil}

\def\d{{\rm d}}
\def\e{{\rm e}}

\numberwithin{equation}{section}

\newtheorem{teorema}{Theorem}
\newtheorem{corollario}{Corollary}

\def\phi{{\varphi}}
\newcommand{\equal}{\buildrel {\rm def} \over {=} }

\newcommand{\dif}{\mathrm{d}}
\newcommand{\blank}{\vskip 1ex \noindent} 
\newcommand{\mfasi}{\mathcal{M}}
\newcommand{\spconfig}{\Gamma}
\newcommand{\supporto}{\mathrm{supp}}

\begin{document}

\title{An averaging theorem for FPU in the thermodynamic limit}
\author{A.~Maiocchi, D.~Bambusi, A.~Carati}

\date{\today}
\maketitle

\begin{abstract}
Consider an FPU chain composed of $N\gg 1$ particles, and endow the
phase space with the Gibbs measure corresponding to a small
temperature $\beta^{-1}$. Given a fixed $K<N$, we construct $K$
packets of normal modes whose energies are adiabatic invariants (i.e.,
are approximately constant for times of order $\beta^{1-a}$,
$a>0$) for initial data in a set of large measure. Furthermore,
the time autocorrelation function of the energy of each packet does
not decay significantly for times of order $\beta$.  The restrictions
on the shape of the packets are very mild. All estimates are uniform
in the number $N$ of particles and thus hold in the thermodynamic
limit $N\to\infty$, $\beta>0$.
\end{abstract}

\section{Introduction}

In 1954 Fermi, Pasta and Ulam, being interested in the problem of the
foundations of statistical mechanics, started the study of the energy
exchanges among the normal modes of a nonlinear chain of particles with
nearest neighbor interaction. In the present paper we prove a result
bounding the variation of the energy of packets of normal modes for
times of the order $\beta^{1-a}$, with $a>0$, where
$\beta>0$ is the inverse temperature of the chain. The bound holds
for initial data in a set of large Gibbs measure.  We also prove that
the time autocorrelation function of each packet remains
significantly away from zero at least for times of order $\beta$. As
far as we know this is the first rigorous result on energy exchange
among packets of modes of the FPU model in the thermodynamic limit.

\vskip5pt

The FPU model has been the object of a huge number of studies (see
e.g. \cite{GalFPU} for a report and \cite{GPP12,CGGP07,BCP13} for some
numerical works strictly related to the present one), and many
techniques have been used in order to give significant analytical
predictions about the dynamics of the chain. We recall in particular
the averaging type results of \cite{SW00,BP}, the results on the
dynamics of solitary waves of
\cite{FPI,FPII,FPIII,FPIV,HW08,Miz09,Miz13}, and the results of
\cite{BKP} on the Toda chain. However all known results cover only the
case of small {\it total} energy, so that they are unable to deal with
the thermodynamic limit (with finite \emph{specific} energy) which is
the relevant one for foundations of statistical mechanics.

A technique allowing one  to deal with the thermodynamic limit was
introduced in \cite{C,CM} (see also \cite{RH13,GPP13}); this is the
technique that we extend here to deal with the FPU system. We recall
that the idea of those papers was to consider a ``resonant'' linear
combination $\Phi_0:=\sum_k\nu_kI_k$ of the actions $I_k$ of
the linearized system and to construct a modification
$\Phi:=\Phi_0+\Phi_1$ whose Poisson bracket with the Hamiltonian has
a zero of high  order at the origin. Then one uses methods from
statistical mechanics in order to estimate the ratio between the
standard deviation of $\dot \Phi$ and that of $\Phi$. Finally one can
use standard probabilistic techniques to deduce the result on the
variation in time of $\Phi$ (and of $\Phi_0$), and of
their time autocorrelation functions.

In order to apply such ideas to the FPU system we have to tackle two
kinds of difficulties, which we think should appear also in typical
models of crystal dynamics. The first one is related to
the fact that low temperature FPU is a perturbation of a linear system
presenting a {\it continuum of frequencies}, so that the problem of
small denominators (which was absent in \cite{CM}) occurs here in a
new way.\footnote{Small denominators appear also in \cite{C,RH13},
  where however the frequencies occur essentially as iid random
  variables.} This problem
is here overcome exploiting two properties, the first one is that, due
to the translational invariance of the interactions, there occurs a
selection of the coefficients actually appearing in the interaction,
which in turn implies a selection rule on the small denominators. The
second property is that if one stops the construction at order three,
then the small denominators always appear with a numerator which
depends on the coefficients $\nu_k$ defining $\Phi_0$. Thus, with an
appropriate choice of $\nu$, the numerators are made to vanish exactly
when the the denominators do. Surprisingly enough, such a procedure only
imposes a constraint on the behavior of $\nu_k$ as $k\to 0$ (see Theorem
\ref{teor:funzionale} below) and thus one has a great freedom in the
choice of the adiabatic invariants. The fact that at order four more
complicated small denominators appear constitutes an obstruction to a naive
extension of the present result to longer time scales.

The second difficulty tackled here is related to the fact that the
normal modes of the unperturbed system (linearized FPU) are the Fourier
modes, while the measure presents in a simple way if it is written in
the space of the particles. So we have to work quite a lot in order to
perform, in an efficient way, the averages of the
quantities of interest.

\vskip 10pt

The paper is organized as follows: in Sect. \ref{sez:risultati} we
give a precise statement of our results; in Sect. \ref{sez:principale}
we prove the result on the adiabatic invariance of the energies of
packets of normal modes; such a section is split into two subsections:
in the first one we give the proof of the main theorem using the
result of the main technical Lemma \ref{lemma:stima} which is proved
in the subsequent subsection. In Sect. \ref{sez:funzionale} we prove
Theorem \ref{teor:funzionale} which gives a simple characterization of
the allowed functions $\nu$. Finally, in the Appendix \ref{app:misura}
we giva the proof of a more or less standard auxiliary Lemma useful
for the computation of averages.

\noindent {\it Acknowledgments.} During the preparation of this work
we had many very interesting discussions with the colleagues of the
groups of Milano and Padova (G. Benettin, L. Galgani, A. Giorgilli,
S. Paleari, T.Penati, A. Ponno) that we warmly thank. This research
was founded by the Prin project 2010-2011 ``Teorie geometriche e
analitiche dei sistemi Hamiltoniani in dimensioni finite e infinite''.


\section{Stability estimate for the FPU model}\label{sez:risultati}
The Hamiltonian of the FPU--system with fixed end points can be
written, in suitably rescaled variables, as 
\begin{equation}
\label{H}
{H=H_0+H_1+H_2}
\end{equation}
where
\begin{eqnarray*}
H_0&\equal &\sum_{j=0}^N\left(\frac{p_j^2}2+\frac{\left(q_{j+1}-
  q_j\right)^2}2
\right)\ ,\\
H_1&\equal& \frac 13\sum_{j=0}^N\left(
q_{j+1}- q_j\right)^3\\
H_2&\equal& \frac A4\sum_{j=0}^N \left( q_{j+1}- q_j\right)^4\ ,
\end{eqnarray*}
and $p=(p_1,\ldots,p_N)$, $q=(q_1,\ldots,q_N)$ are canonically
conjugated variables in the phase space $\mfasi\equiv\R^{2N}$,
$p_0=p_{N+1}=q_0=q_{N+1}=0$, and $A>0$ is a positive parameter.

We endow the phase space by the Gibbs measure at inverse
temperature $\beta$, namely
\begin{equation}
\label{gibbs}
\d \mu(p,q)\equal \frac{\e^{-\beta H(p,q)}}{Z(\beta)} \d^{n}p\d^{n}q\ ;
\end{equation}
as usual $Z(\beta)$ is the partition function, i.e. the normalization
constant such that the measure of $\mfasi$ equals 1. Given a function
$F$ on the phase space, we will use this 
measure to compute its average $\langle F\rangle$, its $L^2$-norm
$\norma{F}$ and its variance $\sigma^2_{F}$ defined by
\begin{align}
\label{ave}
&\langle F\rangle\equal \int_{\mfasi} F\d \mu\ ,
\\
\label{norma}
& \norma{F}^2\equal \int _{\mfasi} |F|^2\d \mu\ ,
\\
\label{norma1}
& \sigma_F^2\equal \norma{F-\langle F\rangle }^2\ .
\end{align}

We define also the correlation of two dynamical variables $F,G$ by
$$
C_{F,G}:={\langle FG\rangle -\langle F\rangle\langle
  G\rangle} 
$$
and the time autocorrelation of a dynamical variable by 
\begin{equation}
\label{norma2}
C_F(t):=C_{F,F(t)}\ ,
\end{equation}
where $F(t):=F\circ g^t$ and $g^t$ is the flow of the FPU system. 

 The unperturbed Hamiltonian $H_0$ can be put in diagonal form by
 passing to the normal modes of oscillation. The canonically
 conjugated coordinates of the normal modes, denoted by $\hat p=(\hat
 p_1, \ldots, \hat p_N)$ and $\hat q=(\hat q_1,\ldots, \hat q_N)$ are
 obtained through the canonical change of variables
\begin{equation*}
\begin{split}
p_j= \sqrt{\frac{2}{N+1}}\sum_{k=1}^N \hat p_k \sin\left(\frac{\pi
  j k}{N+1} \right)\ , \\
q_j= \sqrt{\frac{2}{N+1}}
\sum_{k=1}^N \hat q_k \sin\left(\frac{\pi j k}{N+1} 
\right) \ .
\end{split}
\end{equation*}
In such variables, $H_0$ takes the form
$$
H_0= \sum_{k=1}^{N} \frac{\hat p_k^2}2 +\frac{\omega_k^2 \hat
  q_k^2}{2}= \sum_{k=1}^{N} \omega_k I_k\ ,
$$
where we have defined the actions
$$
I_k\equal \frac{\hat p_k^2+\omega_k^2 \hat q_k^2}  {2\omega_k} 
$$
and the frequencies $\omega_k= 2\sin\left(\frac{\pi
  k}{2(N+1)}\right)$.
Thus the FPU system at low temperature turns out to be a small
perturbation of $H_0$, the perturbation parameter being
$\beta^{-1/2}$. 

Let $\nu\in \mathcal C^1([0,1],\R^+)$ be a differentiable function;  
as anticipated above, we are interested in the time evolution of
quantities of the form 
 $$\Phi_0\equal
\sum_{k=1}^N \nu\left(\frac{k}{N+1}\right) I_k\ .
$$ 
In the following we will often denote $\nu_k\equal\nu(k/(N+1))$;
furthermore we define $\omega(x):=2\sin(\pi x/2)$ so that 
$\omega(k/(N+1))=\omega_k$. 

Theorem~\ref{teor:principale} below controls the time variation of (a
small perturbation of) $\Phi_0$ in terms of the functional
$h(\nu)\equal (h_1(\nu)+1)/h_2(\nu)$, defined by
\begin{eqnarray}
 h_1(\nu)&\equal&\max_{\tau_i=\pm 1}\sup_{x,y\in[0,1]}
\left|\frac{\tau_1\nu(x)+ \tau_2\nu(y)+\tau_3
  \nu(z(x,y))}{\tau_1\omega(x)+
  \tau_2\omega(y)+ \tau_3\omega(z(x,y))} \right|\ , \label{eq:definizione_h_1}\\
h_2(\nu)&\equal & \int_0^1
\frac{\nu^2(x)}{\omega^2(x)}\dif\, x\ ,
\label{eq:definizione_h_2}
\end{eqnarray}
$$
z=z(x,y)\equal\left\{
\begin{array}{cc}
 x + y &\mbox{if }x+y\le 1\\
 2-x-y &\mbox{if }x+y> 1
\end{array}
\right.\ .
$$


Our main result is the following theorem, which will be proved in the
rest of the paper.
\begin{teorema}\label{teor:principale}
Let $\nu(x)$ be such that $h_1(\nu)<\infty$
and $g(x):=\nu(x)/\omega(x)$ has bounded derivative. Define
$\Phi_0\equal \sum_k \nu(k/(N+1)) I_k$, then there exist constants
$\beta^*>0$, $ N^*>0$ and $C>0$ s.t., for any $\beta>\beta^*$ and for
any $N> N^*$, there exists a polynomial of third order $\Phi_1$ with
the property that $\Phi\equal \Phi_0+\Phi_1$ fulfills
\begin{equation}\label{eq:principale}
\frac{\|\dot{\Phi}\|}{\sigma_{\Phi}}\le \frac{C}\beta h(\nu)
\ .
\end{equation}
\begin{equation}\label{eq:vicinanza}
\sigma_{\Phi_1}/\sigma_{\Phi_0}<Ch(\nu)/\sqrt{\beta}\ .
\end{equation}
\end{teorema}

\blank \textbf{Remark 2.} The theorem is almost
void if one cannot estimate the quantity $h(\nu)$ as a functional of
$\nu$. Whereas the denominator $h_2(\nu)$ is simply related to the
fraction of energy contained in the packet, it is more complicated to
have an estimate of the numerator $h_1(\nu)$. However, under some
regularity assumption on $\nu$, an upper bound to $h_1(\nu)$ is
provided in terms of the supremum of $g(x)\equal \nu(x)/\omega(x)$ and
of its second derivative by the following theorem, whose proof can be
found in Section~\ref{sez:funzionale}.
\begin{teorema}\label{teor:funzionale}
Let $\nu(x)$ be such that $g(x)\in\mathcal C^2([0,1],\R)$ and $g'(0)=
0$, and set $c_0\equal g(0)$, $c_2\equal\sup_{x\in[0,1]}|
g''(x)|$. Then there exists a constant $C>0$, independent of $\nu$,
such that one has
$$
h_1(\nu)\le C (c_0+c_2)\ .
$$
Moreover, if $g'(0)\neq 0$, $h(\nu)$ is not bounded.
\end{teorema}

It is worth to point out some consequences of the main theorem:

\begin{corollario}\label{cor:correlazioni}
In the hypotheses of Theorem~\ref{teor:principale}, there exists $C_1$
s.t. 
\begin{equation}
\frac{\mathbf{C}_{\Phi_0}(t)}{\sigma^2_{\Phi_0}}\ge
\frac{1}{2}\ ,\quad \forall |t|\leq\frac{\beta}{C_1} .
\end{equation}
\end{corollario}
\proof One starts by observing that, in virtue of Theorem~3 of
\cite{CM}, (\ref{eq:principale}) implies that
\begin{equation}\label{eq:aggiunta}
\mathbf{C}_\Phi(t) \ge \sigma^2_\Phi \left(1-\frac{C^2
  h^2(\nu)}{2\beta^2} t^2\right)\ ,
\end{equation}
whereas, applying Schwartz inequality one gets
\begin{equation*}
\left| \sigma^2_\Phi-\sigma_{\Phi_0}^2\right| = \left|
\sigma^2_{\Phi_1} +2\langle (\Phi_1-\langle \Phi_1\rangle
);(\Phi_0-\langle \Phi_0\rangle)\right| \le \sigma^2_{\Phi_1}+
2\sigma_{\Phi_1} \sigma_{\Phi_0}
\ .
\end{equation*}

On the other hand (cf. also Theorem~1 of \cite{correlazioni}) one also has
$$
\left|\mathbf{C}_\Phi(t)-\mathbf{C}_{\Phi_0}(t)\right| = \left|
\mathbf{C}_{\Phi_1} (t)\right|+ 2\left|\langle \Phi_0;
\Phi_1\circ g^t\rangle \right| \le
\sigma^2_{\Phi_1}+2\sigma_{\Phi_1}\sigma_{\Phi_0}  \ .
$$
Since (\ref{eq:vicinanza}) provides the upper bound
\begin{equation}\label{eq:aggiunta_2}
 \sigma^2_{\Phi_1}+2\sigma_{\Phi_1}\sigma_{\Phi_0}  \le
 \sigma^2_{\Phi_0} \left(\frac{C^2 h^2(\nu)}{ \beta}+ 2 \frac{C 
  h(\nu)}{\sqrt \beta}\right)\ ,
\end{equation}
the thesis then follows.\qed

We have also the following corollary on the probability $\mathbf P$
that the time evolution of $\Phi_0$ is large:
\begin{corollario}
In the hypotheses of Theorem~\ref{teor:principale}, there exists $C_2$
s.t. $\forall\, 0\leq a\leq 1/2$ one has
\begin{equation}
\label{prob.1}
\mathbf P\left(|\Phi_0(t)-\Phi_0|\ge \frac{
  \sigma_{\Phi_0}}{\beta^{a/2}}\right) 
\le \frac{C_2}{\beta^{a}}\ ,\quad \forall |t|\leq\beta^{1-a}   \ ,
\end{equation}
where, as above, $\Phi_0(t)=\Phi_0\circ g^t$.
\end{corollario}
\proof The proof is easily done by using the relations
\begin{equation}\label{eq:varianza_variazione}
\sigma^2_{\Phi_0(t)-\Phi_0}= 2\left(\sigma^2_{\Phi_0}- \mathbf{C}_{\Phi_0
}(t)\right) \le 2\sigma^2_{\Phi_0}\left( \frac{C_1h^2(\nu)}{\sqrt \beta}+
\frac{C_1 h^2(\nu)}{2\beta^2} t^2\right)\ ,
\end{equation}
where in the upper bound use is made of
\eqref{eq:aggiunta}, \eqref{eq:aggiunta_2}. Then one applies the Chebyshev
inequality to $\Phi_0(t)-\Phi_0$, which gives, for any
$\lambda>0$:
\begin{align*}
\mathbf P\left(|\Phi_0(t)-\Phi_0|\ge \lambda
  \sigma_{\Phi_0}\right) 
= \mathbf P\left(|\Phi_0(t)-\Phi_0|\ge 
  \frac{\lambda \sigma_{\Phi_0}}{\sigma_{\Phi_0(t)-\Phi_0}}
  \sigma_{\Phi_0(t)-\Phi_0}\right) 
\\
\le 
  \frac{\sigma^2_{\Phi_0(t)-\Phi_0}} {\lambda^2 \sigma^2_{\Phi_0}}\ .
\end{align*}
By choosing $\lambda= \beta^{-a/2}$ and inserting relation
(\ref{eq:varianza_variazione}) the thesis is proved.
\qed

\begin{remark}
\label{rem.gio}
Following \cite{GPP12} it is also possible to bound the probability that
the time average and the time variance of $\Phi_0(t)-\Phi_0$ is not
small. Here, for simplicity we choose to state just the previous Corollary.
\end{remark}

Of course one can repeat the argument for different choices of the
function $\nu$. In particular, having fixed an integer $K$ independent
of $N$, one can define $K$ different functions $\nu^{(1)},
\nu^{(2)},..., \nu^{(K)} $, for example with disjoint support, each one
fulfilling the assumptions of Theorem \ref{teor:principale}, so that
the quantities $\Phi_0^{(l)}\equal \sum_k \nu_k^{(l)}I_k $ are
adiabatic invariants. Precisely one has
\begin{corollario}
Assume that $\nu^{(l)}$, $l=1,...,K$ fulfill the assuptions of Theorem~\ref{teor:principale}, there exists $C_3$
s.t. $\forall\, 0\leq a\leq 1/2$ one has
\begin{equation}
\label{prob.12}
\mathbf P\left(\exists l\ :\ |\Phi^{(l)}_0(t)-\Phi^{(l)}_0|\ge \frac{
  \sigma_{\Phi_0^{(l)}}}{\beta^{a/2}}\right) 
\le \frac{C_3}{\beta^{a}}\ ,\quad \forall |t|\leq\beta^{1-a}   \ .
\end{equation}
\end{corollario}

\section{Proof of Theorem~\ref{teor:principale}}\label{sez:principale}

\subsection{The proof}
In this section we give the proof of Theorem~\ref{teor:principale}
using the results of the main technical Lemma \ref{lemma:stima}, which
will be proved in the subsequent subsection.

The proof consists in performing the first step of the formal
construction of an integral of motion which is a perturbation of
$\Phi_0$, and in estimating its time derivative. Define $
\Phi=\Phi_0+\Phi_1$, with $\Phi_1$ a polynomial of order three
determined by the condition that $\{\Phi,H\}$ is of order four, where $\{.,.\}$ denotes the Poisson bracket. Then
$\Phi_1$ must fulfill the equation 
\begin{equation}\label{eq:omologica}
\{H_0,\Phi_1\}=-\{H_1,\Phi_0\}\ .
\end{equation}
The formal construction is standard (see, for instance
\cite{giorgilli}), but the estimate of the remainder requires a
special care and is the main difficulty we have to address here.

To start with we pass to 
the complex coordinates
$$
\xi_k=\frac{\hat p_k+i\omega_k\hat q_k}{\sqrt2}\ , \eta_k=\frac{\hat
  p_k-i\omega_k\hat q_k}{\sqrt2} \ ,
$$
such that $\{\xi_k,\eta_k\}=i\omega_k$ and $H_0=\sum_k \xi_k
\eta_k$. 
Then the nonlinearity is a linear combination of monomials of the form 
$$
\Xi^s_{\tau,k} \equal\xi_{k_1}^{(1+\tau_1)/2 }
\eta_{k_1}^{(1 -\tau_1)/2}\ldots 
\xi_{k_s}^{(1+\tau_s)/2}\eta_{k_s}^{(1-\tau_s)/2} \ ,\quad s\geq 3
$$
where
\begin{equation}
\label{indici.1}
\tau=(\tau_1,...,\tau_s)\ ,\ \tau_l=\pm1\ ,\quad
k=(k_1,...,k_s)\ ,\ k_l=1,...,N\ ;
\end{equation}
furthermore, the
index $k$ is such that
\begin{equation}
\label{sel}
\mod{\tilde \tau\cdot k}= 0\ ,\quad\text{where}\ \mod{n}\equal
n\ \text{mod}[ 2(N+1)]\ ,
\end{equation}
for some
\begin{equation}
\label{indici3}
\tilde \tau=(\tilde \tau_1,...,\tilde \tau_s)\ ,\quad \tilde
\tau_l=\pm1\ .
\end{equation}
In the following we will use denote by $\I_s$ the set of the indexes
$(\tau,\tilde \tau,k)$ of the form \eqref{indici.1},
\eqref{indici3}. Finally, for $i\in\Z$ we will denote
$$
\delta_i=\left\{ 
\begin{matrix}
1 & \mbox{if}\ i=0
\\
0 & \mbox{otherwise}
\end{matrix}
\right.\ .
$$

\begin{definition}
\label{def.P}
We say that $f\in \mathcal P_s$ if it has the form 
\begin{equation}
\label{ps}
f=\frac 1{(N+1)^{(s-2)/2}}\sum_{(\tau,\tilde \tau,k)\in \I_s}
f_{\tau,\tilde \tau}\left(\frac{k_1}{N+1},\ldots,\frac{k_s}{N+1}\right)
\Xi^s_{\tau,k} \delta_{\mod{ \tilde \tau\cdot k}}\ ,
\end{equation}
where $f_{\tau}:[0,1]^s\to\C$ are continuous functions.
\end{definition}

This is the
class of polynomials which will enter the perturbative construction.

We define in $\mathcal P_s$ the norm
\begin{equation}
\label{norma.f}
\left\|f\right\|_+\equal \max_{(\tau,\tilde \tau, k)\in \I_s}
\left|f_{\tau,\tilde \tau}\left(\frac{k_1}{N+1},\ldots,\frac{k_s}{N+1}
\right)\right|\delta_{\mod{\tilde \tau\cdot k}}\ .
\end{equation}

The variance of a dynamical variable in $\mathcal P_s$ is related to
the above defined norm by the following lemma which is the main
technical lemma of the paper and whose proof is deferred to
subsection \ref{app:stima}. 
\begin{lemma}\label{lemma:stima} For any integer $s\geq 2$
  there exist $N_0>0$ and $C$ such that, for any $N>N_0$, and any 
$f\in\mathcal P_s$ one has
\begin{equation*}
\sigma^2_f\le N \frac{C}{\beta^{s}} \left\| f\right\|^2_+\ .
\end{equation*}
\end{lemma}
The norm of the Poisson brackets of two variables is
controlled by the following lemma whose simple proof is omitted.
\begin{lemma}\label{lemma:par_poisson}
 If $f\in \mathcal P_s$, $g\in
  \mathcal P_r$, then $\{f,g\}\in\mathcal P_{r+s-2}$. Moreover, one has
$$
\left\|\{f,g\}\right\|_+ \le 2^4 \max(s,r) \left\|f\right\|_+
\left\|g\right\|_+\ .
$$
\end{lemma}

In order to find a solution of equation (\ref{eq:omologica}), we
express $H_1$ in complex coordinates, namely
\begin{equation*}
\begin{split}
H_1&= \frac {i}6\sqrt{\frac1{N+1}}\sum_{k_1,k_2,k_3=1}^N \left(
\xi_{k_1}-\eta_{k_1}\right) \left( \xi_{k_2}-\eta_{k_2}\right) \left(
\xi_{k_3}-\eta_{k_3}\right)\\ &\times
\left(3\delta_{k_1+k_2-k_3}+\delta_{k_1+k_2+k_3- 2(N+1)}\right)
\end{split}
\end{equation*}
so that $H_1\in \mathcal P_3$ (one can similarly check that $H_2\in
\mathcal P_4$). Then, by using the properties of Poisson brackets and
the fact that $\Phi_0= \sum_k (\nu_k/\omega_k) \xi_k \eta_k$, one can
check that a formal solution of (\ref{eq:omologica}) is given 
by the expression

\begin{equation*}
\begin{split}
\Phi_1= \frac {i}3\sqrt{\frac1{N+1}}\sum_{{\tau_i=\pm 1}\atop{k_i=1,...,N}} &
\tau_1 \tau_2\tau_3
    \frac{\tau_1 \nu_{k_1}+ \tau_2 \nu_{k_2}
  +\tau_3\nu_{k_3}}{ \tau_1 \omega_{k_1}+
  \tau_2 \omega_{k_2}  +\tau_3\omega_{k_3}}
    \Xi^3_{\tau,k} \\
&\times \left(3\delta_{k_1+k_2-k_3}+\delta_{k_1+k_2+k_3-
  2(N+1)}\right)\ .
\end{split}
\end{equation*}
Clearly $\Phi_1$ is well defined if $h_1(\nu)$ is bounded.

\noindent {\it Proof of Theorem \ref{teor:principale}.}  We
 bound  the numerator of the fraction at the l.h.s. of
(\ref{eq:principale}) by using Lemma~\ref{lemma:par_poisson} and
Lemma~\ref{lemma:stima} (notice that
$H_0,\Phi_0\in \mathcal P_2$):
$$
\|\dot {  \Phi}\| = \|\{\Phi_1,H_1+H_2\}+\{\Phi_0,H_2\}\|\le \sqrt{N}
\frac{C_0}{\beta^2} (h_1(\nu)+1)\ ,
$$ for some $C_0>0$. Concerning the denominator of
\eqref{eq:principale}, we write
\begin{equation}\label{eq:diff_varianze}
\sigma_{\Phi}\ge \sigma_{\Phi_0}-\sigma_{\Phi_1}
\end{equation}
and we estimate $\sigma_{\Phi_0}$ from below using $\sigma_{\Phi_0}\ge
\sigma_F$ with $F\equal \sum_k (\nu_k/\omega_k) \hat p_k^2/2$,
where the last inequality is due to the stochastic
independence of $\hat p_k$ and $\hat q_k$. Thus one has 
\begin{equation}\label{eq:sigma_F}
\sigma_F^2= \frac{1}{2\beta} \sum_k \left(\frac{\nu_k}{\omega_k}
\right)^2\ge \frac N{4\beta} h^2_2(\nu)\ ,
\end{equation}
where the last estimate  is obtained through Euler summation formula,
which in turn can be applied in virtue of the regularity hypotheses on
$\nu(x)/\omega(x)$. Moreover, notice that, because of the same
hypotheses, $h_2(\nu)$ is bounded from below, so that $h_1(\nu)<\infty$
implies that $h(\nu)<\infty$. On the other hand, one can apply
Lemma~\ref{lemma:stima} and get 
$$
\sigma_{\Phi_1}\le \sqrt{N}
\frac{C_1}{\beta^{3/2}} h_1(\nu)\ ,
$$
for some $C_1>0$. This, together with (\ref{eq:sigma_F}), proves
formula (\ref{eq:vicinanza}). Furthermore, making use again of
(\ref{eq:sigma_F}) and inserting it in (\ref{eq:diff_varianze}),
formula (\ref{eq:principale}) is proved too.
\qed

\subsection{Proof of Lemma~\ref{lemma:stima}}\label{app:stima}
The proof consists in some steps, the first of which is the choice of
suitable coordinates in which the integrals with respect to
Gibbs measure become tractable. The rest of the proof consists of a
careful analysis of the expression obtained through the integration. 

Concerning the choice of coordinates, first we go back to the
variables $\hat p,$ $\hat q$, then the integration over the $\hat p'$s
is easy (they are iid Gaussian variables with zero average).
The integration with respect to the $\hat q$ variables is more
complicated. In order to do it we use the fact that the Hamiltonian is
a simple function of $r_j\equal q_{j+1}-q_j$, for $j=0,\ldots, N$. In
fact, the potential part of the Hamiltonian can be written as
$$
\sum_{j=0}^N V(r_j)\, \quad\mbox{with } V(r)\equal \frac{r^2}2+
\frac 13 r^3+\frac A4 r^4 \ ,
$$ so that the configurational part of the probability measure is
factorized in terms of the variables $r_j$, which are independently
distributed, apart from the constraint $\sum_j r_j= q_{N+1}-q_0=0$
(this implies that they are exchangeable random variables as defined
e.g. in \cite{scambiabili}). The situation is similar to that of the
microcanonical ensemble for the perfect gas, in which the energies of
the particles are independently distributed, except for the constraint
that their sum is fixed. In such a case one can compute mean values
and variances of sensible observables in the canonical ensemble, in
which all energies are independent, and then estimate the error
introduced. For this reason, we will use the mixed coordinates $\hat
p$, $r$, and adopt the methods developed in the frame of statistical
mechanics to deal with the integration over the $r$'s (see
\cite{chincin}). The corresponding lemma\footnote{In its statement, we adopt the
  multi--index notation: $k=(k_0,\ldots, k_N)$ and
  $j=(j_0,\ldots,j_N)$ are vectors of nonnegative integers, with the
  norm defined by $|k|=k_0+\ldots+k_N$. So, $r^k= r_0^{k_0}\cdot
  \ldots \cdot r_N^{k_N}$. Moreover, $\supporto\, k$ denotes the set
  of sites $i$ for which $k_i\neq 0$.} (see Lemma \ref{lemma:misura} below)  is more or less
standard, however, we were not able to find an adapted statement in
literature,so we give its proof in Appendix~\ref{app:misura}.

\begin{lemma}\label{lemma:misura}
There exist $K,N_0>0$ such that, for any multi--index $k,j$ with
length $n$ and $m$, respectively, and any $N>N_0$, one has
\begin{equation}\label{eq:varianza_stessi_siti}
\left|\langle r^k r^l\rangle -\langle r^k\rangle \langle r^l\rangle\right| \le
 K^{n+m} \sqrt{n! m!} \beta^{-(n+m)/2}  \ .
\end{equation}
Moreover, if the sets of sites $\supporto\, k$ and $\supporto\, l$ are
disjoint, one has
\begin{equation}\label{siti_diversi}
\left|\langle r^k r^l\rangle -\langle r^k\rangle \langle r^l\rangle\right| \le
\frac 1N  K^{n+m} \sqrt{n! m!} \beta^{-(n+m)/2}\ .
\end{equation}
\end{lemma}

The variance of $f\in \mathcal P_s$, can be written as
\begin{equation}\label{eq:varianza_estesa}
\begin{split}
\sigma^2_f=&\frac1{(N+1)^{s-2}}
\sum_{{(\tau,\tilde \tau ,k)\in\I_s\atop (\tau',\tilde \tau\null' ,k')\in\I_s}}
f_{\tau,\tilde \tau}\left(\frac{k_1}{N+1},\ldots,\frac{k_s}{N+1}\right)
f_{\tau',\tilde \tau'}\left(\frac{k'_1}{N+1},\ldots,\frac{k'_s}{N+1}\right) \\
&\times\left(\langle \Xi^s_{\tau,k}
\Xi^s_{\tau',k'}\rangle - \langle
\Xi^s_{\tau,k}\rangle \langle\Xi^s_{\tau',k'}\rangle
\right)
\\
&\times\delta_{\mod{\tilde\tau\cdot k}  }
\delta_{\mod{\tilde\tau\null'\cdot k'} }
\end{split}
\end{equation}

Introducing the coordinates $\hat p$, $\hat q$, each term of the second line of
(\ref{eq:varianza_estesa}) gives rise to at most $2^{2s}$ terms of the
form 
\begin{align}
\label{rp}
&\langle \hat r_{k_1}...\hat r_{k_{s_1}}\hat r_{k'_1}...\hat
r_{k'_{s'_1}}\rangle \langle\hat p_{k_{s_1+1}}...\hat
p_{k_{s}}\hat p_{k'_{s'_1+1}}...\hat p_{k'_{s}}\rangle
\\
\nonumber
&-
\langle \hat r_{k_1}...\hat r_{k_{s_1}}\rangle\langle \hat r_{k'_1}...\hat
r_{k'_{s'_1}}\rangle \langle\hat p_{k_{s_1+1}}...\hat
p_{k_{s}}\rangle\langle\hat p_{k'_{s'_1+1}}...\hat p_{k'_{s}}\rangle
\ ,
\end{align}
where $\hat r_k\equal\omega_k\hat q_k$.

The main step of the proof consists in computing a representation
formula for the quantity
\begin{equation}\label{eq:varianza_estesa_2}
\hat A_{k,k'}\equal a\langle \hat r_{k_1}\cdots \hat r_{k_{s_1}} \hat
r_{k'_1}\cdots \hat r_{k'_{s_1'}}\rangle - b\langle \hat r_{k_1}\cdots
\hat r_{k_{s_1}} \rangle\langle\hat r_{k'_1}\cdots \hat
r_{k'_{s_1'}}\rangle \ ,
\end{equation}
where $a,b$ are complex
constants.

We start by establishing some notation. We will denote
\begin{align*}
S\equal s_1+s_1'\ ,\quad
L\equiv(L_1,...,L_S)\equal(l_1,...,l_{s_1},l'_{1}....,l'_{s'_1}) 
\\
K\equiv(K_1,...,K_S)\equal(k_1,...,k_{s_1},k'_{1}....,k'_{s'_1}) \ .
\end{align*}
Inserting the definition of the 
Fourier coefficients, one has
\begin{equation}
\label{Fou.1}
\hat r_k=\sqrt{\frac2{N+1}}\sum_{l=0}^Nr_l\cos\left[\frac \pi
  {N+1}\left(l+\frac12\right)k\right]\ ,
\end{equation}
in \eqref{eq:varianza_estesa_2} one gets
\begin{align}
\nonumber
&\hat A_{k,k'}\equiv \hat A_{K}=\frac{2^{S/2}}{(N+1)^{S/2}}
\\
\label{fou.2}
&\times \sum_{L_1,...,L_S} A_L  \cos\left[\frac 
\pi {N+1}\left(L_1+\frac12\right)K_1\right]\ldots \cos\left[\frac 
\pi {N+1}\left(L_S+\frac12\right)K_S\right]
\end{align}
where 
\begin{equation}
\label{cl}
A_L\equal  a\langle r_{l_1}\cdots r_{l_{s_1}}  r_{l'_1}\cdots 
 r_{l'_{s'_1}}\rangle - b\langle r_{l_1}\cdots r_{l_{s_1}} \rangle\langle
r_{l'_1}\cdots  
 r_{l'_{s'_1}}\rangle\ .
\end{equation}
In order to compute $\hat A_{K}$ we proceed by reducing iteratively
the number of variables to be summed. We will start by summing over
$L_S$. At each step one gets that the quantity to be summed is the
linear combination of quantities of the form \eqref{fou.2} with
coefficients enjoying a suitable property which is the same fulfilled
by averages of exchangeable variables. 

Precisely, let $S$ be an integer and we consider the sequences
$B_{L_1,...,L_S}$ of complex numbers with the property that , if one
fixes $S-1$ indexes, say $L_1,...,L_{S-1}$, then $ B_{L_1,...,L_{S}}$
has the same value for all values of the remaining index, say $L_{S}$,
s.t.
$$
L_S\not=L_1\ \text{and}\ L_S\not=L_2\ \text{and}\ ...\ \text{and}\ L_S\not=L_{S-1}\ . 
$$

\begin{definition}
\label{ts}
We will denote by $\tilde B_{K_1,...,K_S}\equiv \tilde B_K$ the
rescaled Fourier transform of one of these sequences, precisely
\begin{align}
\label{ts.def}
 \tilde B_K &\equal
\sum_{L_1,...,L_{S-S_1}} B_{L_1,...,L_{S-S_1} }
\\
\nonumber
&\times
\cos\left[\frac \pi
  {N+1}\left(L_1+\frac12\right)K_1\right]\ldots \cos\left[\frac \pi
  {N+1}\left(L_{S-S_1}+\frac12\right)K_S\right]
\end{align}
where $ B_{L_1,...,L_{S-S_1}}$ has the property just described. 
\end{definition}

The main remark needed in order to start the induction is contained in
the following Lemma.

\begin{lemma}
\label{lemmapoi.1}
The following formula holds:
\begin{align}
\nonumber
&\tilde B_K
=(N+1)\delta_{\mod{K_S}} \widetilde {B^0}\null_{K_1,K_2,...,K_{S-1}} 
\\
\nonumber
&+\widetilde {B^1}\null_{K_1+K_S,K_2,...,K_{S-1}} 
+\widetilde {B^1}\null_{K_1-K_S,K_2,...,K_{S-1}}
\\
\label{ts.ind.1}
&+...+ \widetilde {B^{S-1}}\null_{K_1,K_2,...,K_{S-1}+K_S}
+ \widetilde {B^{S-1}}\null_{K_1,K_2,...,K_{S-1}-K_S}\ ,
\end{align}
where the $\widetilde{B^j}$ are obtains through \eqref{ts.def} from
the sequences
\begin{align}
\label{b0}
B^{0}_{L_1,...,L_{S-1}}&\equal B_{L_1...L_S}\big|_{L_S\not=L_1,...,L_S\not=L_{S-1}}\ , 
\\
\label{bj}
B^{j}_{L_1,...,L_{S-1}}&\equal\frac{
B_{L_1,...,L_{S-1},L_j}- B_{L_1...L_S}\big|_{L_S\not=L_1,...,L_S\not=L_{S-1}}}{2}\ . 
\end{align}
\end{lemma} 
\proof It is a computation which exploits the formula  
\begin{equation}\label{eq:delta_coseni}
\sum_{L=0}^N\cos\left[\frac 
\pi {N+1}\left(L+\frac12\right)k\right]=(N+1)\delta_{\mod{k} }\ .
\end{equation}
In order to use it we rewrite $\tilde B_K$ by separating the sum over
$L_S$, namely
\begin{align*}
\tilde B_K=\sum_{L_1,...,L_{S-1}}\cos\left[\frac \pi
  {N+1}\left(L_1+\frac12\right)K_1\right]\ldots \cos\left[\frac \pi
  {N+1}\left(L_{S-1}+\frac12\right)K_{S-1}\right] \\ \times \left\{
B_{L_1...L_S}\big|_{L_S\not=L_1,...,L_S\not=L_{S-1}} \sum_{L_s:
  L_S\not=L_1,...,L_S\not=L_{S-1}} \cos\left[\frac \pi
  {N+1}\left(L_{S}+\frac12\right)K_{S}\right]\right.
\\ +B_{L_1,...,L_{S-1},L_1} \cos\left[\frac \pi
  {N+1}\left(L_{1}+\frac12\right)K_{S}\right]+...
\\
\left.+B_{L_1,...,L_{S-1},L_{S-1}}
\cos\left[\frac \pi {N+1}\left(L_{S-1}+\frac12\right)K_{S}\right]
\right\}
\end{align*}
but the curly bracket is equal to 
\begin{align*}
&B_{L}\big|_{L_S\not=L_1,...,L_S\not=L_{S-1}} \sum_{L_S=0}^N \cos\left[\frac \pi
  {N+1}\left(L_{S}+\frac12\right)K_{S}\right]
\\ 
&+\left(B_{L_1,...,L_{S-1},L_1}-B_{L}\big|_{L_S\not=L_1,...,L_S\not=L_{S-1}} \right) \cos\left[\frac \pi
  {N+1}\left(L_{1}+\frac12\right)K_{S}\right]+...
\\
&+\left(B_{L_1,...,L_{S-1},L_{S-1}}-B_{L}\big|_{L_S\not=L_1,...,L_S\not=L_{S-1}}
\right)  
\cos\left[\frac \pi {N+1}\left(L_{S-1}+\frac12\right)K_{S}\right]
\\
&=B_{L}\big|_{L_S\not=L_1,...,L_S\not=L_{S-1}}\delta_{\mod{K_s}}(N+1)+
\sum_{j=1}^{S-1}B^j_{L_1,...,L_{S-1}} \cos\left[\frac \pi
  {N+1}\left(L_{j}+\frac12\right)K_{S}\right]
\end{align*}
where 
\begin{equation}
\label{eqb.1}
B^j_{L_1,...,L_{S-1}} \equal
B_{L_1,...,L_{S-1},L_{j}}-B_{L}\big|_{L_S\not=L_1,...,L_S\not=L_{S-1}}
. 
\end{equation}
In order to conclude the proof insert such a formula in the expression
for $\tilde B_K$ and remark that except for the term containing the $\delta$,
all the other addenda contain the expression
\begin{align*}
\cos\left[\frac \pi
  {N+1}\left(L_{j}+\frac12\right)K_{S}\right]\cos\left[\frac \pi
  {N+1}\left(L_{j}+\frac12\right)K_{j}\right]=
\\
 \frac{1}{2}\left\{ \cos\left[\frac \pi
  {N+1}\left(L_{j}+\frac12\right)(K_{S}+K_j)\right]+ \cos\left[\frac \pi
  {N+1}\left(L_{j}+\frac12\right)(K_{S}-K_j)\right] \right\}
\end{align*}
so that the thesis follows. \qed

With formula \eqref{ts.ind.1} at hand we can iterate the
construction in order to get the general structure of the terms
involving $\hat A_{K}$. 

Actually, in order to get the proof of Lemma \ref{lemma:stima}, we
need quite precise information on the structure of
$\hat A_{K}$. To this end we still need some more preliminary definitions.

Having fixed a positive integer $S$, we consider vectors ${\mathbf{\tau}}\equiv
(\tau_1,...,\tau_S)$, with $\tau_j\in \{-1,0,1\}$. The set of such
vectors will often be denoted by $\Z_3^S$. 

\begin{definition}
\label{admi}
A collection $\tau^{(1)},...,\tau^{(S_1)}$, $S_1\leq S$ of vectors of
$\tau^{(i)}\in\Z^S_3$ will be said to be $S$-admissible, or simply
admissible, if the following properties hold
\begin{itemize}
\item[1)] the supports supp$(\tau^{(i)})$ are disjoint.
\item[2)] $\bigcup_{i=1}^{S_1}$supp$(\tau^{(i)})=\{1,...,S\}$. 
\end{itemize} 
\end{definition}

%
We are now ready for the main lemma of this section. It gives the
representation formula for the $A$'s. 

\begin{lemma}
\label{main.esti}
$\hat A_{K}$ is the sum of a number independent of $N$ of
addenda, each one of the form
\begin{equation}
\label{btau}
B_{{\mathbf{\tau}}}\left[ \prod_{i=1}^{S_1} \delta_{\mod{\tau^{(i)}\cdot K}}
  \right] (N+1)^{S_1-S/2}\ ,\quad S_1\leq S
\end{equation} 
where ${\mathbf{\tau}}=(\tau^{(1)},...,\tau^{(S_1)})$ is an $S$-
admissible collection of vectors.

Furthermore, $B_{\mathbf{\tau}}$ is a linear combination of the
quantities $A_L$ (cf. eq. \eqref{cl}), such that the indexes $L$
assume only those values s.t. the following property holds
\begin{equation}
\label{selection}
\left[I\in{\rm supp}(\tau^{(i)})\ ,\ J\in{\rm supp}(\tau^{(j)})\ , i\not=j
  \right]\Longrightarrow L_I\not=L_J\ .
\end{equation}
The number of terms in the linear combination is bounded independently
of $N$, the coefficients are bounded uniformly with respect to $N$.
\end{lemma}
\proof The proof is obtained by applying iteratively Lemma
\ref{lemmapoi.1}. We claim that, after $R$ steps of decomposition,
$\tilde B_K$ turns out to be the sum of terms of the form 
\begin{align}
\label{SR}
\tilde B_{K\cdot\tau^{(1)},...,K\cdot\tau^{(S-R)}}
\left[
(N+1)^{S_1}\prod_{i=1}^{S_1} \delta_{\mod{K\cdot\tau^{(S-R+i)}}}
\right]\ ,
\\
\nonumber
\text{with}\quad S_1\leq R\leq S\ ,
\end{align}
where ${\mathbf{\tau}}\equiv(\tau^{(1)},...,\tau^{(S+S_1-R)})$ is an admissible
collection and the $B$'s fulfill a variant of the selection property
\eqref{selection}. Precisely, define $$\bar {\mathbf{\tau}}\equal
(\tau^{(1)}+...+\tau^{(S-R)},\tau^{(S-R+1)},...,\tau^{(S-R+S_1)}),$$
then $B$ fulfills \eqref{selection} with respect to such a collection
of vectors (which is $S$-admissible). 

We prove \eqref{SR} by induction on $R$. The
formula is true for $R=0$ with $S_1=0$. We assume it is true for $R$
and we prove it for $R+1$. 

Applying \eqref{ts.ind.1} to \eqref{SR}, such a quantity turns out to
be the sum of 
\begin{equation}
\label{s+0}
\widetilde{B^0}_{K\cdot\tau^{(1)},...,K\cdot\tau^{(S-R-1)}}
\left[
(N+1)^{S_1+1}\prod_{i=1}^{S_1+1} \delta_{\mod{K\cdot\tau^{(S-R+i)}}}
\right]\ ,
\end{equation}
and of the quantities
\begin{equation}
\label{SR.1}
\widetilde{B^j}_{K\cdot\tau^{(1)},...,K\cdot\tau^{(j)}+K\cdot\tau^{S-R},...,
 K\cdot\tau^{(S-R-1)}}
\left[
(N+1)^{S_1}\prod_{i=1}^{S_1} \delta_{\mod{K\cdot\tau^{(S-R+i)}}}
\right]\ ,
\end{equation}
so also at step $R+1$ we have the wanted representation. Still we have
to verify that the new ${\mathbf{\tau}}$'s form an admissible collection and that
the involved coefficients $A_L$ fulfill the selection rule
\eqref{selection}.  

We start by the $\tau$'s in the term \eqref{s+0}. In this term the
collection of the ${\mathbf{\tau}}$'s coincides with the previous one,
so it is still an admissible collection.

We come to the selection rule on $B^0$. The new collection $\bar
{\mathbf{\tau}}$ is
$$(\tau^{(1)}+...+\tau^{(S-R-1)},\tau^{(S-R+1)},...,\tau^{(S-R-1+S_1)})\ ,$$
so that one is adding a further restriction on the values of the
indexes of the $A_L$'s entering in $B^0$, namely that $L_{S-R}$ (and
thus also the indexes labeled by $\supp (\tau^{(S-R)})$) must be
different from the other indexes. But, by formula \eqref{b0} one has
$$
B^0_{L_1,...,L_{S-R-1}}=B_{L}\big|_{L_{S-R}\not=L_1,...,
  L_{S-R}\not=L_{S-R-1}}\ ,
$$
and therefore the selection rule is fulfilled.

We come to \eqref{SR.1}. In this term the elements of the collection of
the ${\mathbf{\tau}}$'s are the same as before except for the fact that
$\tau^{S-R} $ is missing and that $\tau^{(j)}$ is substituted by 
$$
\tau^{(j)\pm}\equal \tau^{(j)}\pm\tau^{(S-R)}\ .
$$
By the properties of the supports one has thus
\begin{equation}
\label{supp}
\supp (\tau^{(j)\pm})=\supp(\tau^{(j)})\cup \supp(\tau^{(S-R)})\ ,
\end{equation}
from which one immediately sees that properties 1) and 2) of
definition \ref{admi} are fulfilled also by the new collection. 

Concerning the selection property for $B$, we just notice that the new
collection $\bar {\mathbf{\tau}}$ coincides with the old one, and therefore the
new $B$'s automatically fulfill the needed property. \qed

We have now to insert the averages of the  $\hat p$'s. To get a
useful formula we have to analyze quite in detail the corresponding
terms. 

First remark that
a possible expression of
$\langle\hat p_{k_1}...\hat p_{k_s}\rangle$ is constructed as follows:
consider  the distinct partitions of $1,...,s$ into subsets
composed by an even number of elements. Let $\Sigma^s\equiv
(\Sigma^s_1,...,\Sigma^s_{s_1})$ be one of these partitions (of course
$s_1\leq s/2$), denote $\ell_J\equal\#\Sigma^s_J$, and let 
$j^{(J)}_1,...,j^{(J)}_{\ell_J}\in\Sigma^s_J$ be the elements of
$\Sigma^s_J$, then to the partition $\Sigma^s$ we associate the quantity
\begin{equation}
\label{sigma}
D_{\Sigma^s}\equal \left[\prod_{J=1}^{s_1}\langle\hat
  p_1^{\ell_J}\rangle
  \right]2^{s/2}\beta^{s/2}=\prod_{J=1}^{s_1} (\ell_J-1)!!
\end{equation}
and one has
\begin{equation}
\label{pks}
\langle\hat p_{k_1}...\hat p_{k_s}\rangle=\frac{1}{2^{s/2}\beta^{s/2}}
\sum_{\Sigma^s} D_{\Sigma^s}\delta_{k_1,...,k_s}^{\Sigma^s}\ ,
\end{equation}
\begin{equation*}
\delta_{k_1,...,k_s}^{\Sigma^s}\equal 
\left\{   
\begin{matrix}
1 & \text{if}\ k_{j_1^{(J)}}=...=k_{j_{\ell_J}^{(J)}}\ \forall J
\\
0 &\text{otherwise}  
\end{matrix}
\right. \ ,
\end{equation*}
where, of course the sum is over all the distinct partitions described
above. 

\begin{remark}
\label{ks}
Defining ${\bf k}^{(J)}\equiv (k^{(J)}_1,...,k^{(J)}_s)$ with
$k^{(J)}_{j^{(J)}_i}=1$ for $i=1,...,\ell_J$ and zero otherwise, one
has 
\begin{equation}
\label{deltaEk}
\delta_{{\bf k}}^\Sigma\not=0\ \iff\ {\bf k}=\sum_{J=1}^{s_1}n_J{\bf
  k}^{(J)}\ ,
\end{equation}
for some integers $n_J$. This means that for every fixed partition
$\Sigma$, the subspace of vectors in $\mathbb Z^{s}$ such that
$\delta_{{\bf k}}^\Sigma\not=0$ has dimension $s_1\le s/2$, where the
equality is attained only if $\ell_J=2$, $\forall J$.
\end{remark}

For this reason the partitions for which $\ell_J=2$ for all $J$'s will
play a special role. In such a case one can write
$$
\delta^{\Sigma^s}_k=\prod_{i=1}^{s/2}\delta_{k\cdot \tau^{(i)}} ,
$$ where $\tau=\{\tau^{(i)}\}_{i=1}^{s/2}$ is an $s$-admissible
collections s.t. each of the $\tau^{(i)}$'s has only one component
equal to 1 and one component equal to $-1$.
 
{\sl We will denote by $\Tc^s$ the set of the $s$-admissible collections
with such a property. 

We will denote by $\Sc^s_4$ the set of partitions $\Sigma^s$ such that
$\ell_J\geq 4$ for at least one $J$.}

In order to obtain a useful expression for the covariance we consider
$\Tc^{s+s'}$ and decompose it as 
\begin{equation}
\label{ts2}
\Tc^{s+s'}=\Tc^s\oplus\Tc^{s'}=\Tc^s\cup\Tc^{s'}\cup\Tc^{s,s'}\ ,
\end{equation}
where $\Tc^{s,s'}$ is composed by the $(s+s')$-admissible collections
s.t. at least one of the vectors $\tau^{(i)}$ has one non vanishing
component in the $\Tc^s$ and one nonvanishing component in $\Tc^{s'}$.

\begin{lemma}
\label{f.p.l}
The following formula holds
\begin{align*}
a \langle\hat
p_{k_{1}}\cdots \hat p_{k_s} \hat p_{k'_{1}}\cdots \hat p_{k'_{s'}}
\rangle -b\langle\hat
p_{k_{1}}\cdots \hat p_{k_s} 
 \rangle\langle \hat p_{k'_{1}}\cdots
\hat p_{k'_{s'}} \rangle 
\\
= \frac{1}{(2\beta)^{\frac{s+s'}{2}}}\left[
\sum_{{\tau\in\Tc^{s}\atop \tau'\in\Tc^{s'}}}(a-b)\left(\prod_{
  i=1}^{s/2}\delta_{k\cdot\tau^{(i)}} \right)
\left(\prod_{ i=1}^{s'/2}\delta_{k\cdot\tau^{'(i)}} \right)\right.
\\
\left.+a\sum_{\tau\in\Tc^{s,s'}}\prod_{
  i=1}^{(s+s')/2}\delta_{K\cdot\tau^{(i)}}+
\sum_{\Sigma^{s+s'}\in\Sc^{s+s'}_4}E_{\Sigma^{s+s'}}\delta_{K}^{\Sigma^{s+s'}}
\right]
\end{align*}
where $K=(k,k')$ and $E_{\Sigma^{s+s'}}$ is a (possibly vanishing)
constant fulfilling
$$
|E_{\Sigma^{s+s'}}|\leq C(|a|+|b|).
$$
\end{lemma} 

The proof is a simple computation which is omitted.

We have now at hand the tools that enable us to estimate $\sigma_f$.
In the forthcoming formulas we will use the following notations:
$S_1\leq s_1+s_1'$ is an integer and
\begin{align*}
k=(k_1,...,k_s)\ ,\quad k'=(k'_1,...,k'_s)
\\
k^{(1)}=(k_1,...,k_{s_1})\ ,\quad k^{(2)}=(k_{s_1+1},...,k_{s})
\\
k^{'(1)}=(k'_1,...,k'_{s'_1})\ ,\quad k^{'(2)}=(k'_{s_1'+1},...,k'_{s})
\\
K=(k_1,...,k_s,k'_1,...,k'_s)\ ,
\\
K^{(1)}=(k_1,...,k_{s_1},k'_1,...,k'_{s'_1})\ ,\quad
K^{(2)}=(k_{s_1+1},...,k_{s},k'_{s_1'+1},...,k'_{s}) \ ,
\end{align*}
finally $s_2:=s-s_1$ and $s'_2:=s-s'_1$.

First remark that, due to Lemma~\ref{main.esti} and \eqref{pks} one
has that $\sigma_f^2$ is estimated by the sum of finitely many terms
of the form 
\begin{align}
\label{1.B}
\frac{C}{(N+1)^{s-2}}\norma f_+^2
\sum_{(k,k')\in\Z^{2s}}\delta_{\mod{\tilde\tau\cdot k}   } 
\delta_{\mod{\tilde\tau\null'\cdot k'} }
\\
\label{1.D}
\left[\langle\hat
p_{k_{s_1+1}}\cdots \hat p_{k_s} \hat p_{k'_{s'_1+1}}\cdots \hat p_{k'_s}
\rangle \langle r_{l_1}....r_{l_{s_1}}r_{l'_1}...r_{l_{s_1}'}\rangle
  \right.
\\
\label{1.E}
\left.
-\langle\hat
p_{k_{s_1+1}}\cdots \hat p_{k_s} 
 \rangle\langle \hat p_{k'_{s'_1+1}}\cdots
\hat p_{k'_s} \rangle
\langle r_{l_1}....r_{l_{s_1}}\rangle\langle r_{l'_1}...r_{l_{s_1}'}\rangle
\right]
\\
\label{1.F}
\times \left(\prod_{i=1}^{S_1}\delta_{\mod{\tau^{(i)}\cdot
    K^{(1)}}}(N+1)\right) \frac{1}{(N+1)^{(s_1+s_1')/2}}
\end{align}
where $l_1,...,l_{s_1},l'_1,...,l'_{s'_1}$ fulfills the selection rule
\eqref{selection} with respect to the partition $\tau^{(i)}$.  

According to Lemma \ref{f.p.l} one has that \eqref{1.B}-\eqref{1.F}
can be written as
$$
\Sigma_1+\Sigma_2+\Sigma_3\ ,
$$
where
\begin{align}
\label{2.1.B}
\Sigma_1:=\frac{C}{(N+1)^{s-2}}\norma f_+^2
\sum_{(k,k')\in\Z^{2s}}\delta_{\mod{\tilde\tau\cdot k}  } 
\delta_{\mod{\tilde\tau\null'\cdot k'} }\frac{1}{(2\beta)^{\frac{s_2+s_2'}{2}}}
\\
\label{2.1.D}
\sum_{{\tau'\in\Tc^{s_2}\atop \tau''\in\Tc^{s'_2}}}\left(
\langle r_{l_1}....r_{l_{s_1}}r_{l'_1}...r_{l_{s_1}'}\rangle-\langle
r_{l_1}....r_{l_{s_1}}\rangle\langle r_{l'_1}...r_{l_{s_1}'}\rangle 
\right)\left(\prod_{
  i=1}^{s_2/2}\delta_{k^{(2)}\cdot\tau'\null^{(i)}} \right)
\left(\prod_{ i=1}^{s'_2/2}\delta_{k^{(2)}\cdot\tau^{''}\null^{(i)}} \right)
\\
\label{2.1.F}
\times \left(\prod_{i=1}^{S_1}\delta_{\mod{\tau^{(i)}\cdot
    K^{(1)}}}(N+1)\right) \frac{1}{(N+1)^{(s_1+s_1')/2}}
\end{align}

\begin{align}
\label{2.2.B}
\Sigma_2:=\frac{C}{(N+1)^{s-2}}\norma f_+^2
\sum_{(k,k')\in\Z^{2s}}\delta_{\mod{\tilde\tau\cdot k}  } 
\delta_{\mod{\tilde\tau\null'\cdot k'} }\frac{1}{(2\beta)^{\frac{s_2+s_2'}{2}}}
\\
\label{2.2.C}  \langle r_{l_1}....r_{l_{s_1}}r_{l'_1}...r_{l_{s_1}'}\rangle
  \sum_{\tau'\in\Tc^{s,s'}}\prod_{
  i=1}^{(s_2+s_2')/2}\delta_{K^{(2)}\cdot\tau'\null^{(i)}}
\\
\label{2.2.F}
\times \left(\prod_{i=1}^{S_1}\delta_{\mod{\tau'\null^{(i)}\cdot
    K^{(1)}}}(N+1)\right) \frac{1}{(N+1)^{(s_1+s_1')/2}}
\end{align}

\begin{align}
\label{2.3.B}
\Sigma_3:=\frac{C}{(N+1)^{s-2}}\norma f_+^2
\sum_{(k,k')\in\Z^{2s}}\delta_{\mod{\tilde\tau\cdot k}  } 
\delta_{\mod{\tilde\tau\null'\cdot k'} }\frac{1}{(2\beta)^{\frac{s_2+s_2'}{2}}}
\\
\label{2.3.C}
\sum_{\Sigma^{s_2+s_2'}\in\Sc^{s_2+s_2'}_4}E_{\Sigma^{s_2+s_2'}}\delta_{K^{(2)}}^{\Sigma^{s_2+s_2'}}
\\
\label{2.3.F}
\times \left(\prod_{i=1}^{S_1}\delta_{\mod{\tau^{(i)}\cdot
    K^{(1)}}}(N+1)\right) \frac{1}{(N+1)^{(s_1+s_1')/2}}
\end{align}
where the indexes $l_1,...,l_{s_1},l'_1,...,l'_{s'_1}$ fulfills the
selection rule \eqref{selection} with respect to the collection
$\tau$.

\begin{lemma}
\label{+di4}
The following estimate holds
\begin{equation}
\label{+4}
\left|\Sigma_3\right|\leq\frac{C(N+1)\norma f_+^2}{\beta^s}
\end{equation}
\end{lemma}
\proof For this computation we can neglect the delta's in
\eqref{2.3.B}. Every $\delta$ in \eqref{2.3.F} reduces by 1 the
effective dimension of the lattice over which $K^{(1)}$ runs. Thus,
the effective dimension of such a lattice is $s_1+s_1'-S_1\geq 0$. By
remark \ref{ks}, $K^{(2)}$ runs over a lattice of dimension at most
$(s-s_1+s-s_1')/2-1$. Thus, the number of nonvanishing terms is at
most of order
$$
(N+1)^{\wedge}\left( s+\frac{s_1+s'_1}{2}-S_1-1 \right) ,
$$
while, counting the powers of $(N+1)$, one has that each term has size
controlled by a constant times 
$$
\frac{\norma
  f_+^2}{\beta^s}(N+1)^{\wedge}\left(S_1-\frac{s_1+s_1'}{2}-s+2\right)  ,
$$
so that the result follows. \qed

We have now to understand when it can happen that the deltas coming
from the zero momentum conditions are not independent of the other
deltas (more precisely the corresponding $\tau$ vectors). This
is analyzed by the forthcoming Lemma \ref{ind.tau}.

Write $\Z^{2s}=\Z^s\oplus \Z^s$ and denote by $P_1$ the projection on
the first factor and by $P_2$ the projection on the second one. Then
the following Lemma holds.

\begin{lemma}
\label{ind.tau}
Let $\tau^{(i)}$ be a $(2s)$-admissible collection of vectors and let
$\tilde \tau\in\Z^{2s}_3$ be a vector with support equal to $(1,...,s)$, namely
s.t. $\tilde \tau_i\not=0$ $\forall i=1,...,s$ and $\tilde \tau_i=0$
$\forall i=s+1,...,2s$. Assume that there exists $\bar \imath$
s.t. $P_1\tau^{(\bar\imath)}\not=0 $ and  $P_2\tau^{(\bar\imath)}\not=0 $,
then $\tilde \tau$ is linear independent of the vectors $\tau^{(i)}$.
\end{lemma}
\proof Consider the equation 
$$
c\tilde \tau+\sum_ic_i\tau^{(i)}=0 \ ;
$$
applying $P_1$ and $P_2$ one gets
\begin{align}
\label{t.1}
cP_1\tilde \tau+\sum_ic_iP_1\tau^{(i)}=0 \ ,
\\
\label{t.2}
\sum_ic_iP_2\tau^{(i)}=0 \ .
\end{align}
Since the supports of the $\tau^{(i)}$'s are disjoint, \eqref{t.2}
implies $c_i=0$ for all $i$'s s.t. $P_2\tau^{(i)}\not=0$. In
particular one has $c_{\bar\imath}=0$. There exists a component of
$P_1\tau^{(\bar\imath)}$ which is different from zero. Assume for
definiteness that it is the first one. It follows that all the other
vectors $\tau^{(i)}$ have first component equal to zero. Thus, taking
the first component of \eqref{t.1} one gets
$$
c\tilde \tau_1+c_{\bar\imath}\tau^{(\bar i)}_1=c\tilde
\tau_1=0\ \Longrightarrow c=0\ ,
$$ 
which is the claimed independence.
\qed

In particular it follows that, in the expression of $\Sigma_2$, at
least one of the $\tilde\tau$'s is independent of all the other
$\tau$'s. Thus the Following Lemma holds

\begin{lemma}
\label{indip}
The following estimate holds
\begin{equation}
\label{sigma2}
\left|\Sigma_2\right|\leq\frac{C(N+1)\norma f_+^2}{\beta^s}
\end{equation}
\end{lemma}
\proof Every $\delta$ reduces by 1 the
effective dimension of the lattice over which $K$ runs, provided the
corresponding vectors $\tau$ are independent. In the considered case
the effective dimension is at most 
$$
2s-(S_1+s-(s_1+s'_1)/2) ,
$$
thus, counting the powers of $(N+1)$ as in the proof of Lemma
\ref{+di4} one gets the result. \qed

To estimate $\Sigma_1$ one has also to consider the dependent
case. This is contained in the proof of the following Lemma.

\begin{lemma}
\label{dip1}
The following estimate holds
\begin{equation}
\label{sigma1}
\left|\Sigma_1\right|\leq\frac{C(N+1)\norma f_+^2}{\beta^s}
\end{equation}
\end{lemma}
\proof The case in which the $\tau's$ are independent is dealt with as
in the proof of Lemma \ref{indip}. Consider now the case in which they
are dependent. In such a case, all the elements of $\tau$ do not mix
$k$ and $k'$, by the selection rule \eqref{selection}, it follows that
the indexes $l_1,...,l_{s_1}$ are all different of the indexes
$l'_1,...,l'_{s_1'}$, thus the covariance in \eqref{2.1.D} can be
estimated using eq. \eqref{siti_diversi}, which adds a power of $N$ at
the denominator. Thus the result follows also in this case. 
\qed


\section{Proof of Theorem~\ref{teor:funzionale}}\label{sez:funzionale}
As both $\nu$ and $\omega$ are bounded from above, $h_1(\nu)$ can
diverge only when the denominator at the r.h.s. of
(\ref{eq:definizione_h_1}) vanishes. We prove that, under the
assumptions of the theorem, the numerator vanishes at the same points
and the ratio stays bounded.

First, remark that the hypotheses on $g$ and the explicit form of
$\omega(x)=2\sin(\pi x/2)$ imply that $\nu$ has derivative bounded by
$K\equal 2c_2+\pi(c_0+ c_2/2)$. In turn, this implies also that the
numerator is bounded by $3K$.

Consider now the case in which $z=x+y$. When $\tau_1=\tau_2=\tau_3$,
the inequality $\sin(\pi x/2)\ge x$ implies that the denominator is
bigger than $3x$. Using $\nu(x)\le Kx$ one has that the ratio which defines
$h_1$ is bounded by $K$. 

When $\tau_1=-\tau_2=\tau_3$, using the fact that $\omega$ is a
non--decreasing function one can bound the denominator from below by
$x$; in turn, using $|\nu(x+y)-\nu(y)|\le Kx$ one has that the
numerator is smaller than $2 K x$, so that the ratio is smaller than
$2K$.  The same upper bound holds when $-\tau_1=\tau_2=\tau_3$.

The case $\tau_1=\tau_2=-\tau_3$ is more complicated. We rewrite
the numerator of (\ref{eq:definizione_h_1}) as a function of $g(x)$
and get
\begin{equation*}
\begin{split}
\nu(x)+\nu(y)-\nu(x+y)= & c_0 \left[\omega(x)+\omega(y)-\omega(x+y)
  \right] \\
& +f(x)+f(y)-f(x+y)
\end{split}
\end{equation*}
where we have put $f(x)\equal \omega(x) (g(x)-c_0)$. Due to its
definition and to the hypothesis $g'(0)=0$, $f(x)$ has a zero of third
order at 0. Suppose, without loss of generality, that
$y\le x$, then there exists a constant $C$,
such that one has
$$
|f(x)-f(x+y)|\le C c_2 x^2y\ ,\quad |f(y)|\le C c_2 y^3\le C c_2 x^2y\ .
$$
On the other hand, for the denominator one has
\begin{equation}\label{eq:minorazione_omega}
\omega(x)+\omega(y)-\omega(x+y)\ge 2\sin(\pi y/2)(1-\cos(\pi
x/2))\ge  x^2 y\ ,
\end{equation}
where in the first inequality use is made of the addition formulas for
the sine, in the second of the inequalities $\sin(\pi
x/2)\ge x$ and $\cos(\pi x/2)\le 1- x^2/2$. Thus we have
$$
\left|\frac{ \nu(x)+\nu(y)-\nu(x+y)}{\omega(x)+\omega(y)
  -\omega(x+y)}\right| \le C (c_0+c_2)\ ,
$$ with a suitable redefinition of the constant $C$. A similar
arguments, exchanging upper with lower bounds, shows that if
$g'(0)\neq 0$ the ratio defining $h_1(\nu)$ is unbounded thus proving
the last statement of the theorem.

Consider now the case of $z=2-x-y$. Here, when $\tau_1=\tau_2=\tau_3$
everything is trivial, because at least one among $x$, $y$ and $z$ is
greater than $2/3$, so that the denominator is larger than
$2\sin(\pi/3)=\sqrt 3$. There remains the case in which one sign is
different from the others: as the role of $x$, $y$ and $z$ is
symmetric, we consider only the possibility
$\tau_1=\tau_2=-\tau_3$. Since $\omega(2-\alpha)=\omega(\alpha)$, one
has
$$
\omega(x)+\omega(y)-\omega(2-x-y)= \omega(x)+\omega(y)-\omega(y+z)\ ,
$$
so that we can bound from below the denominator making use of
inequality (\ref{eq:minorazione_omega}); assuming again without loss
of generality that $y\le x$. In the present case, however, one has
$x+y=2-z \ge 1$, so that $x\ge 1/2$ and we get
$$
\omega(x)+\omega(y)-\omega(2-x-y)\ge \frac{y}{4}\ .
$$
The numerator is bounded according to
$$
|\nu(x)-\nu(2-x-y)|\le K |2(1-x)-y|\le 3 K y\ , \quad |\nu(y)| \le Ky\ ,
$$
where we used the inequality $x\ge 1-y$. This suffices to bound
uniformly the considered ratio also in this case, and thus to complete
the proof.
\qed
\blank

\appendix
\section{Proof of Lemma~\ref{lemma:misura}}\label{app:misura}

As already stated, the main tool needed in the proof of this lemma is
an estimate of the error introduced in computing the mean values of
interest with respect to the measure in which all $r$'s are
stochastically independent, rather than to the one
in which they are conditioned to have vanishing sum. Indeed, if
the $r$'s are independent, estimates (\ref{eq:varianza_stessi_siti}) and
(\ref{siti_diversi}) are trivial consequences of the
properties of Gaussian integration (the r.h.s. of
(\ref{siti_diversi}) even vanishes). The estimates of the
error, as first pointed out by Khinchin (see  
\cite{chincin}), can be obtained by using a local central limit
theorem.

We begin by considering the extended configuration space $\spconfig$,
which coincides with $\R^{N+1}$ endowed with the probability
measure $\mu_{\gamma}$ with density
\begin{equation}\label{eq:densita}
\rho^{N+1}_\gamma (r_0,\ldots,r_N)\equal \frac
1{\left(q_\gamma(\beta)\right)^{N+1}}\prod_{j=0}^N\exp\left(-\gamma
r_j-\beta V(r_j)\right)\ , 
\end{equation}
in which the normalization constant is defined by
$$
q_\gamma(\beta)\equal \int_{-\infty}^\infty \exp\left(-\gamma r-\beta
V(r)\right) \dif r\ .
$$
Denoting by
\begin{equation}
\label{Sigma}
\Sigma_x\equal \left\{ r\in\Gamma\ :\ R\equal \sum_j r_j=x  \right\}\ ,
\end{equation}
the configuration space for our dynamical system corresponds to
$\Sigma_0$. Moreover, the probability measure induced on it by the
Gibbs measure is exactly the measure $\mu_{\gamma}$ with density
(\ref{eq:densita}) with $\gamma=0$ conditioned on $\Sigma_0$. This
means that, if $M$ is a subset of $\Sigma_0$, one has
$$
\mathbf{P}(M)= \mu_0(M|\Sigma_0) \ .
$$
This suggest the introduction of the structure function
$\Omega_{N+1}(x)$, defined by
\begin{equation*}
  \begin{split}
    \Omega_{N+1}(x) &\equal \frac{\dif~}{\dif x} \int_{\sum r_i\le x} e^{-\beta\sum
      V(r_i)} \dif r_0 \ldots\dif r_N = \\
    &=\int  e^{-\beta\sum V(x_i-x_{i-1})} \dif x_1 \ldots\dif x_{N}\ ,
    \quad x_0=0 \ ,\  x_{N+1}=x \ ,
  \end{split}
\end{equation*}
which is the probability density that
$R$ takes on the value $x$ in $\spconfig$, if the $r$'s are
distributed with the measure $\mu_0$. Notice that
$Z(\beta)=\Omega_{N+1}(0)$, which will be used below.

Now the idea is that, being the variable $r_i$ independently
distributed, one can use some kind of central
limit theorem  to expand $\Omega_N$ as a simple
function around zero for large $N$, so that the conditioned measure
becomes easily tractable. 
Indeed, this can be accomplished,
as first pointed out by Cram\`er, if one considers the conjugate
distribution
\begin{equation}\label{eq:distribuzione_coniugata}
U_N^{(\gamma)}(x) = \frac{1}{\Phi_N(\gamma)} e^{-\gamma x} \Omega_N(x)\ ,
\end{equation}
where $\Phi_N(\gamma)=(q_\gamma(\beta))^N $, which correspond
the probability distribution of the sum of independently distributed
random variable distributed  with the density $\rho^N_\gamma$ introduced
above by (\ref{eq:densita}). In fact, while the central limit theorem
gives no direct information on $\Omega_N(x)$, for $x$ near 0 (because
the mean value of $x$ is very far from zero),
such information can
be obtained by applying the central limit theorem to
$U_N^{(\gamma)}(x)$, if one chooses a value of $\gamma$, call it
$\theta$, such that\footnote{Notice that  the integral equation
\begin{equation}\label{eq:equazione_theta}
\begin{split}
0=\int_{-\infty}^{\infty} x U_N^{(\theta)}(x)\, \dif x=\sum_{j=1}^N
\frac1{q_\theta(\beta)}\left( \int_{-\infty}^{+\infty} r
\exp\left(-\theta r -\beta V(r)\right) \dif r\right)\\
\Rightarrow \int_{-\infty}^{+\infty} r
\exp\left(-\theta r -\beta V(r)\right) \dif r=0\ .
\end{split}
\end{equation}
admits a unique solution for all
$\beta>0$ so that $q_\theta(\beta)$ is well defined. It is then
obvious that $\theta$ depends on $\beta$ but not on $N$, so 
that $q_\theta(\beta)$ is a function of $\beta$ only. Furthermore,
since we are interested in the high $\beta$ regime, we point out that
$\lim_{\beta\to \infty} \theta= -\alpha$.}
$$
\int xU_N^{(\theta)}(x)\dif x=0 \ .
$$ 
Then the central limit theorem can be locally applied to
$U_N^{(\theta)}(x)$ near zero and
then translated into a property of $\Omega_N(x)$ by inverting
(\ref{eq:distribuzione_coniugata}).  

We will use the following local version of the central limit, in
which the conjugate distribution is approximated as a function of the
functions $q_j(x)$ defined as
$$
q_j(x)= \frac1{\sqrt{2 \pi}} e^{-x^2/2}\sum H_{j+2s}(x) \prod_{m=1}^j
\frac 1{k_m!} \left(\frac{\gamma_{m+2}}{(m+2)!b^{m+2}}\right)^{k_m}\ , 
$$
where $H_m(x)$ are Hermite polynomials, $\gamma_m$ is the $m$-th
cumulant of $u^{(\theta)}(x)$ and $b$ its standard deviation, while
the sum should be taken on all the non-negative integer solutions
$(k_1,\ldots,k_j)$ of the equalities $k_1+2k_2+\ldots+jk_j=j$, and $s=
k_1+k_2+\ldots k_j$.
\begin{teorema}[Local central limit, Theorem~VII.15 of
    \cite{petrov}]\label{teor:lim_centrale} 
There exist $C,N_0,\beta_0>0$ such that, for $N>N_0,$ $\beta>\beta_0$,
one has
\begin{equation}\label{eq:lim_centrale}
\left|U^{(\theta)}_N(x)- \frac{1}{\sqrt{2\pi N
    b^2}}\exp\left(-\frac{x^2}{2Nb^2}\right)- \sum_{j=1}^2
\frac{q_j(x/(\sqrt N b))}{ N^{(j+1)/2}b}\right| \le \frac{1}{N^{3/2}b}\ ,
\end{equation}
uniformly in $x$.
\end{teorema}
From this theorem we can infer (cf.~\cite{landim}, Corollary~1.4 of
Appendix~2) an estimate on the
deviation of the expectations taken with respect to the Gibbs measure
from that taken with respect to the measure $\mu_\theta\equiv
\mu_\gamma\big|_{\gamma=\theta}$. We denote the expectation of
$f$ with respect to the Gibbs measure by $\langle f\rangle$, while
that with respect to $\mu_\theta$ as $\langle f\rangle_\theta$.
Moreover, given a vector $j=(j_0,\ldots,j_N)\in \{0,1\}^{N+1}$ and a
vector $r\in\Gamma$, we denote by $\tilde r\in\R^{|j|}$ the collection
$r_i$, $i\in\supp j$.

\begin{corollario}\label{cor:landim}
Fix $\bar \beta>0$ and let $f(\tilde r):\mathbb R^{|j|}\to \mathbb R$ have a finite
second order moment with respect to $\mu_\theta$, uniformly for all $\beta>\bar
\beta$. Then there exist $C$, $N_0$ and $\beta_0$ such that, for all
$N>N_0$, $\beta>\beta_0$, one has 
$$
\left|\langle f\rangle -\langle f\rangle_\theta \right|\le C \frac{|j|}{N}
\sqrt{ \langle f^2 \rangle_\theta -\langle f\rangle_\theta^2}
$$
\end{corollario}
\blank
\textbf{Proof.} We denote $J\equal |j|$ and, in order to fix ideas we
assume supp$j=\{ 1,...,J  \}$, the general case is dealt with exactly
in the same way. The average $\langle f\rangle$ can be written as follows
\begin{align*}
\langle f\rangle = \int_{\R^{N+1}} f(\tilde r) \frac{\exp\Big(-\beta\sum
  V(x_i-x_{i-1})\Big)}{Z(\beta)} \dif x_1\ldots\dif x_{N}
\\
=
\int_{\tilde \Gamma} f(\tilde r) \frac{
    \Omega_{N+1-J}(-w)}{\Omega_{N+1}(0)} \dif \tilde v\ ,
\end{align*}

 where $w\equal \sum_{i=0}^Jr_i$, and $\tilde \Gamma\equiv
\R^{N+1-J}$ endowed with the measure with volume element $\dif \tilde
v\equal\prod_{i=0}^J e^{-\beta V(r_i)}\dif r_i$, while $\Omega_{N+1-J}$
is the structure function for the system in which the first $J$
directions are subtracted.

Now the ratio $\Omega_{N+1-J}(-w)/\Omega_{N+1}(0)$ can be expressed
in terms of $U_{N+1}^{(\theta)}(x)$, by a simple inversion of
(\ref{eq:distribuzione_coniugata}), as  
$$
\frac{\Omega_{N+1-J}(-w)}{\Omega_{N+1}(0)}= \frac{
  U^{(\theta)}_{N+1-J}(-w)}{U^{(\theta)}_{N+1}(0)} 
\frac{e^{-\theta w}}{q_{\theta}(\beta)^{J}}\ ,
$$
where  use has been made of the explicit form of
$\Phi_N(\theta)$. So, the difference $|\langle f\rangle -\langle
f\rangle_\theta|$  may be written as
\begin{equation}\label{eq:come_landim}
 \left|
\int_{\tilde \gamma} \dif \tilde v \frac{e^{-\theta
    w}}{q_{\theta}(\beta)^{J}} f(\tilde r)\left(  \frac{
  U^{(\theta)}_{N+1-J}(-w)}{U^{(\theta)}_{N+1}(0)} -1\right)\right|\ .
\end{equation}

Using the relations
$$
\int_{\tilde \gamma} \dif \tilde v \frac{e^{-\theta
    w}}{q_{\theta}(\beta)^{J}}
\frac{U^{(\theta)}_{N+1-J}(-w)}{U^{(\theta)}_{N+1}(0)}
 =  \langle 1 \rangle = 1 = \langle 1 \rangle_\theta=
\int_{\tilde \gamma} \dif \tilde v \frac{e^{-\theta
    w}}{q_{\theta}(\beta)^{J}}
$$
one can rewrite the difference $|\langle f\rangle -\langle
f\rangle_\theta|$ as follows
\begin{equation}\label{eq:come_landim1}
|\langle f\rangle -\langle
f\rangle_\theta|= \left|
\int_{\tilde \gamma} \dif \tilde v \frac{e^{-\theta
    w}}{q_{\theta}(\beta)^{J}} \left(f(\tilde r)- \langle
f\rangle_\theta \right)\left(  \frac{
  U^{(\theta)}_{N+1-J}(-w)}{U^{(\theta)}_{N+1}(0)} -1\right)\right|\ .
\end{equation}

Noting that $|e^{-x^2}-1|\le x^2$ and $q_j(x)\le c_j(\beta_0)$, we
obtain from Theorem~\ref{teor:lim_centrale} that 
$$
\left|  \frac{
  U^{(\theta)}_{N+1-J}(-w)}{U^{(\theta)}_{N+1}(0)} -1\right| \le
K(N_0,\beta_0) \frac   {J}{N} \left(1+ \frac{w^2}{J b^2}
        \right)\ ,
$$
for $N$ large enough. By Schwartz inequality the thesis follows. \qed 
\blank

In order to conclude the proof of Lemma~\ref{lemma:misura} it is now
sufficient to apply corollary~\ref{cor:landim} to the functions of
interest, making use of the properties of Gaussian integration to
estimate $\langle \cdot \rangle_\theta$.

\def\cprime{$'$}

\end{document}